\documentclass{PoS}
\pdfoutput=1
\usepackage[UKenglish]{babel}
\usepackage{natbib}
\newcommand{\gev}{\, {\rm GeV}}
\newcommand{\mev}{\, {\rm MeV}}
\newcommand{\be}{\begin{equation}}
\newcommand{\ee}{\end{equation}}
\newcommand{\bea}{\begin{eqnarray}}
\newcommand{\eea}{\end{eqnarray}}

\title{Pseudoscalar decay constants $f_K/f_\pi$, $f_D$ and $f_{D_s}$ with $N_f = 2 + 1 + 1$ ETMC configurations}

\ShortTitle{Pseudoscalar decay constants $f_K/f_\pi$, $f_D$ and $f_{D_s}$
with $N_f = 2 + 1 + 1$ ETMC configurations}

\author{ P. Dimopoulos$^{(a,b)}$, R. Frezzotti$^{(b,c)}$, P. Lami$^{(d,e)}$, V. Lubicz$^{(d,e)}$, E. Picca$^{(d,e)}$,

 \speaker{L. Riggio}$^{(d,e)}$, G.C. Rossi$^{(b,c)}$, F. Sanfilippo$^{(f)}$, S. Simula$^{(e)}$, C. Tarantino$^{(d,e)}$
\\

\it $^{(a)}$ Centro Fermi - Museo Storico della Fisica e Centro Studi e Ricerche E. Fermi, Rome, Italy

\it $^{(b)}$ Dipartimento di Fisica, Universit\`a di Roma ``Tor Vergata'', Rome, Italy.     E-mail:\email{dimopoulos@roma2.infn.it}, \email{frezzotti@roma2.infn.it}, \email{rossig@roma2.infn.it} %\email{$\{$dimopoulos,frezzotti,rossig$\}$@roma2.infn.it}

\it $^{(c)}$ INFN, Sezione di ``Tor Vergata", Rome, Italy

\it $^{(d)}$ Dipartimento di Matematica e Fisica, Universit\`a  Roma Tre, Rome, Italy.             

Email: \email{lamipaolo@gmail.com}, \email{lubicz@fis.uniroma3.it}, \email{e.picca88@gmail.com} \email{lorenzo.riggio@gmail.com}, \email{tarantino@fis.uniroma3.it}

\it $^{(e)}$ INFN, Sezione di Roma Tre, Rome, Italy. Email: \email{simula@roma3.infn.it}

\it $^{(f)}$ Laboratoire de Physique Th\'eorique (B\^{a}t. 210),
\it Universit\'e Paris Sud, F-91405 Orsay-Cedex, France. Email: \email{fr.sanfilippo@gmail.com}

\\

\bf{For the ETM Collaboration}
}

\abstract{We present a lattice QCD calculation of the pseudoscalar decay constants $f_K$, $f_D$ and $f_{D_s}$ performed by the European Twisted Mass Collaboration with $N_f = 2 + 1 + 1$ dynamical fermions. We simulated at three different values of the lattice spacing, the smallest being approximately $0.06fm$, and with pion masses as small as $210 \mev$. Our main results are: $f_{K^+}/f_{\pi^+}=1.183(17)$, $f_{K^+}=154.4(2.1)\mev$, $f_{D_s}=242.1(8.3)\mev$, $f_D=201.9(8.0)\mev$, $f_{D_s}/f_D=1.199(25)$ and $(f_{D_s}/f_D) / (f_K/f_\pi) = 1.005(15)$.}

\FullConference{31st International Symposium on Lattice Field Theory LATTICE 2013\\
                 July 29 - August 3, 2013\\
                 Mainz, Germany}

\begin{document}

%%%%%%%%%%%%%%%%%%%%%%%%%%%%%%%
\section{Introduction and simulation details}
\label{sec:intro}
%%%%%%%%%%%%%%%%%%%%%%%%%%%%%%%

An accurate determination of the pseudoscalar (PS) meson decay constants is a crucial ingredient for the determination of the CKM matrix elements, and in turn for testing the Standard Model (SM) and searching new physics (NP). 

In this contribution we present a lattice QCD calculation of the $f_K$, $f_D$ and $f_{D_s}$ decay constants using the ensembles of gauge configurations produced by the European Twisted Mass (ETM) Collaboration with four flavors of dynamical quarks ($N_f = 2+1+1$), which include in the sea, besides two light mass degenerate quarks, also the strange and the charm quarks. 
The simulations were carried out at three different values of the inverse bare lattice coupling $\beta$ that allow for a controlled extrapolation to the continuum limit, and at different lattice volumes.
For each ensemble we used a subset of well-separated trajectories to avoid autocorrelations.   
We simulated the pure gauge Iwasaki action \cite{Iwasaki} for gluons, and the Wilson Twisted Mass Action \cite{Frezzotti:2003xj} for sea quarks, which at maximal twist allows for an automatic ${\cal{O}}(a)$-improvement \cite{Frezzotti:2003ni}. 
To avoid mixing in the strange and charm sectors we adopted a non-unitary setup in which valence quarks are simulated for each flavor using the Osterwalder-Seiler action \cite{Osterwalder:1977pc}.
In order to minimize discretization effects in the determination of the pseudoscalar meson masses, the values of the Wilson parameter $r$ are always chosen so that the two valence quarks in a meson have opposite values of $r$.

At each lattice spacing different values of light sea quark masses have been considered. 
The light valence and sea quark masses are always taken to be equal. 
In the light, strange and charm sectors the quark masses were simulated in the ranges $0.1 ~ m_s^{phys} \lesssim  \mu_l \lesssim 0.5 ~ m_s^{phys}$, $0.7 ~ m_s^{phys} \lesssim  \mu_s \lesssim 1.2 ~ m_s^{phys}$ and $0.7 ~ m_c^{phys} \lesssim  \mu_c \lesssim 2.0 ~ m_c^{phys}$, respectively.

We studied the dependence of the PS meson decay constants on the light quark mass fitting simultaneously the data at different lattice spacings and volumes. 
The lattice spacings are: $a = 0.0885(36)$, $0.0815(30)$, $0.0619(18)$ fm \cite{PaoLatt2013}, and the lattice volume goes from $\simeq 2$ to $\simeq 3$ fm. 
The pion masses, extrapolated to the continuum and infinite volume limits, range from $\simeq 210$ to $ \simeq 450 \mev$.

Within our analyses we used: ~ i) the results for $r_0/a$ ($r_0$ is the Sommer parameter \cite{Sommer:1993ce}) obtained from either a linear or a quadratic extrapolation to the chiral limit of the values computed in \cite{Baron:2010bv,Baron:2011sf}, and ~ ii) the quark mass renormalization constant $Z_m = 1/Z_P$ computed in \cite{ETM:2011aa} in the RI-MOM scheme using two different methods, labelled as M1 and M2, which differs by ${\cal{O}}(a^2)$ effects.

\section{Calculation of $f_K$}
In order to compute the kaon decay constant, we first performed a small interpolation of our lattice data to the physical value of the strange quark mass $m_s$ determined in \cite{PaoLatt2013}.
Then we analyzed the dependence of the decay constant as a function of the (renormalized) light-quark mass $m_\ell$ and the lattice spacing and extrapolated it to the physical point.

The SU(2) ChPT prediction at the next-to-leading order (NLO) for $f_K$ reads as follows:
\begin{equation}
\label{eq:fKr0Ch}
f_K = P_1 \left( {1 - \frac{3}{4}\xi _\ell \log \xi _\ell  + P_2 \xi _\ell  + P_3 a^2 } \right) \cdot K^{FSE} _f  ,
\end{equation}
where $\xi_\ell = 2B_0m_\ell / 16\pi^2f_0^2$ with $B_0$ and $f_0$ being the low-energy constants (LECs) entering the LO chiral Lagrangian. The term proportional to $a^2$ in Eq.~(\ref{eq:fKr0Ch}) accounts for leading discretization effects. The factor $K_f^{FSE}$ represents the correction for the finite size effects (FSE) in the kaon decay constant, computed following the work of Colangelo, D\"urr and Haefeli (CDH) \cite{CDH05}.

\begin{figure}[htb!]
\centering
\scalebox{0.25}{\includegraphics{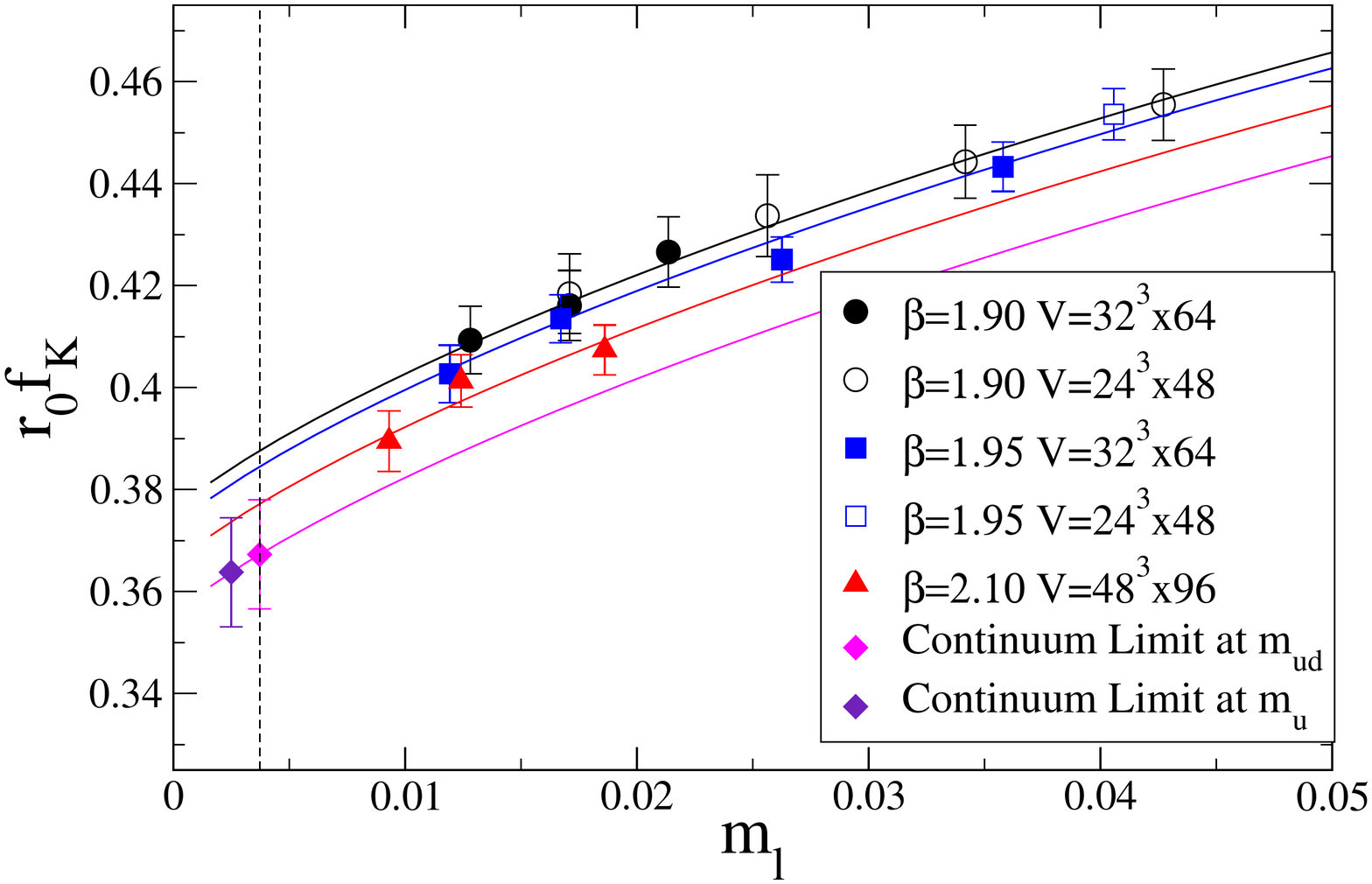}}
\scalebox{0.25}{\includegraphics{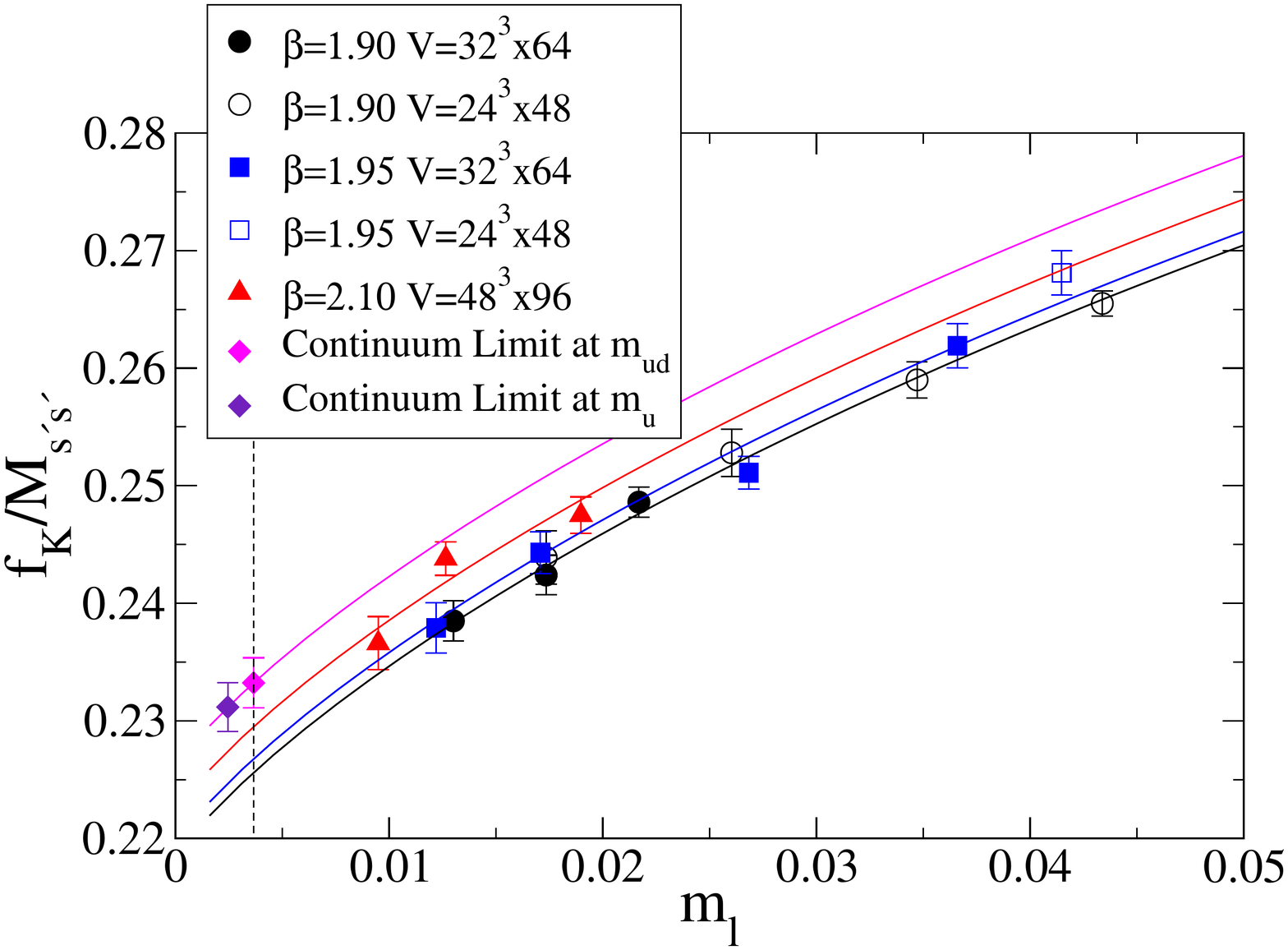}}
\caption{\it Chiral and continuum extrapolation of $r_0 f_K$  (left) and $f_K/M_{s^\prime s^\prime}$ (right) based on the NLO ChPT fit given in Eq.~(\ref{eq:fKr0Ch}). Lattice data have been corrected for FSE using the CDH approach \cite{CDH05}.}
\label{fig:fkchir}
\end{figure}

The chiral and continuum extrapolations of $f_K $ is shown in Fig.~\ref{fig:fkchir} in units of either $r_0$ or the mass $M_{s^\prime s^\prime}$ of a fictitious PS meson made of two strange-like valence quarks with mass fixed at $r_0 m_{s^\prime} = 0.22$.
The impact of discretization effects using $r_0$ as the scaling variable is at the level of $\simeq 3 \%$.
In order to keep the extrapolation to the continuum limit well under control we repeated the analysis adopting a different choice for the scaling variable, namely instead of $r_0$ we introduced the mass $M_{s^\prime s^\prime}$, which is affected by cutoff effects similar to the ones of a K-meson without however any significant dependence on the light quark mass.. 
Thus, we performed the chiral and continuum extrapolation also for the ratio $f_K / M_{s^\prime s^\prime}$.
The comparison between the analyses performed in units of $r_0$ and $M_{s^\prime s^\prime}$ clearly shows that, when $M_{s^\prime s^\prime}$ is chosen as the scaling variable, the discretization effects on $f_K$ change from $\simeq 3 \%$ down to $\simeq - 1.5 \%$.

For the chiral extrapolation we adopted both the NLO ChPT prediction (\ref{eq:fKr0Ch}) and a polynomial formula in $\xi_\ell$. 

Notice in Fig.~\ref{fig:fkchir} that after taking the continuum limit the kaon decay constant has been extrapolated at two different values of the light quark mass. The first one, labelled by a magenta diamond, is the result obtained for $f_K$ in the isospin symmetric limit and corresponds to the extrapolation at the average up/down quark mass $m_{ud} = 3.70 (17) \mev$ \cite{PaoLatt2013}. The second one, represented by a violet diamond, is the value extrapolated to the up quark mass $m_u = 2.47 (11) \mev$, obtained adopting the mass ratio $m_u/m_d=0.5$ from \cite{deDivitiis:2013xla}, and it corresponds to the quantity $f_{K^+}$ corrected for leading strong isospin breaking effect.
In fact, it can be shown that the first order correction due to the QCD isospin breaking effects depends only on the valence quarks. The sea quark effects enter proportionally to the square of the up/down mass difference, $(m_d - m_u)^2$, an effect which is well below the present precision. Thus, in order to correct for leading isospin breaking effects it is sufficient to extrapolate $f_K$ to the up quark mass.
Electromagnetic isospin breaking corrections are a much more challenging issue and have not been considered in the present study.  

The various sources of systematic uncertainties are estimated as follows.
The difference of the results obtained using either $r_0$ or $M_{s^\prime s^\prime}$ as the scaling variable is taken as a systematic uncertainty on $f_K$ associated with discretization effects.
A systematic uncertainty related to the chiral extrapolation is obtained by comparing the results of the NLO ChPT fit with those of the polynomial fit.
As for FSE we compared the results obtained by applying the CDH correction with the ones obtained without correcting for FSE. 
The two methods M1 and M2, used in \cite{ETM:2011aa} to calculate the renormalization constant $Z_P$ in the RI-MOM scheme, allow us to estimate a systematic uncertainty due to the mass renormalization constant.
Finally, the error on our determination of the strange quark mass represents another source of uncertainty, and it has been included in the stat+fit error, which also combines the statistical uncertainty and the error associated with the fitting procedures (dominated by the continuum limit and by the separation between the physical point and the lightest simulated pion mass).
Combining the various sources of uncertainty we find our final result for $f_{K^+}$:
\begin{eqnarray}\label{eq:fk+results}
f_{K^ +} = 154.4(1.8)_{stat + fit} (0.5)_{Chiral} (1.0)_{Disc.} (0.4)_{FSE} (0.2)_{Z_P} \mev =  154.4 (2.1) \mev.
\end{eqnarray}
Dividing the result (\ref{eq:fk+results}) by the experimental value of the pion decay constant, which has been used as input to set the lattice scale \cite{PaoLatt2013}, we obtain for the ratio $f_{K^+ } / f_{\pi^+}$ the value
\begin{eqnarray}\label{eq:fk+/fpi+results}
f_{K^+ } / f_{\pi^+}  = 1.183(14)_{stat + fit} (4)_{Chiral} (8)_{Disc.} (4)_{FSE} (1)_{Z_P} (2)_{f_{\pi^+}} = & 1.183(17) ~ ,
\end{eqnarray}
which can be compared with the FLAG averages $f_{K^+} /f_{\pi^+} = 1.205 (18)$ at $N_f = 2$, $f_{K^+} /f_{\pi^+} = 1.192 (5)$ at $N_f = 2 + 1$ and $f_{K^+} /f_{\pi^+} = 1.195 (5)$ at $N_f = 2 + 1 + 1$ \cite{FLAG}.

In the isospin symmetric limit we get  for $f_K$ the value
\begin{eqnarray}\label{eq:fk}
f_K = 155.6(1.6)_{stat + fit} (0.5)_{Chiral} (1.1)_{Disc.} (0.4)_{FSE} (0.2)_{Z_P} \mev = 155.6 (2.1) \mev ~ ,
\end{eqnarray}
which can be compared with the FLAG averages $f_K = 158.1 (2.5) \mev$ at $N_f = 2$ and $f_K = 156.3 (0.8) \mev$ at $N_f=2+1$ \cite{FLAG}, and for the ratio $f_K / f_\pi$ the result
\begin{eqnarray}\label{eq:fk/fpiresults}
f_K / f_\pi  = 1.193(13)_{stat + fit} (4)_{Chiral} (8)_{Disc.} (4)_{FSE} (1)_{Z_P} (2)_{f_{\pi}} = & 1.193(16) ~ .
\end{eqnarray}
Had we neglected at all discretization effects in the kaon decay constant and had we limited ourselves only to the gauge ensembles at $\beta=1.95$ and $2.10$, the result for $f_K / f_\pi$ would have been larger by $\simeq 2.5 \%$ getting very close to the finding $f_K / f_\pi = 1.224 (13)$ obtained in \cite{Farchioni:2010tb}.

\section{Calculation of $f_{D_s},f_D$ and $f_{D_s}/f_D$}
The physical values of $f_{D_s},f_D$ and $f_{D_s}/f_D$ have been determined by analyzing $\Phi_{D_s} = f_{D_s}\sqrt{M_{D_s}}$ and the double ratio $(f_{D_s}/f_D)/(f_K/f_\pi)$ as functions of $m_c$, $m_s$, $m_\ell$ and $a^2$.

We first performed an interpolation of lattice data for $f_{D_s}, f_D$ and $f_{D_s}/f_D$ to the physical strange and charm quark masses determined in \cite{PaoLatt2013}.
Then the dependence of $\Phi_{D_s}$ on the light-quark mass $m_\ell$ and on the lattice spacing turned out to be described well by the simple expression
\begin{equation}
\label{eq:phids}
\Phi _{D_s} = P_1 (1 + P_2 m_\ell  + P_3 m_\ell ^2  + P_4 a^2  )\,.
\end{equation}
Using the experimental value $M_{D_s} = 1.969 \gev$ allowed us to determine the physical value of $f_{D_s}$.

As in the case of the kaon, the lattice data for $\Phi_{D_s}$ are converted in units of either $r_0$ or the mass $M_{c^\prime s^\prime}$ of a fictitious PS meson, made by one strange-like and one charm-like valence quark with masses fixed at $r_0 m_{s^\prime} = 0.22$ and $r_0 m_{c^\prime} = 2.4$, respectively. 
Such a reference mass $M_{c^\prime s^\prime}$ is expected to have discretization effects closer to the ones of $M_{D_s}$.

\begin{figure}[htb!]
\centering
\scalebox{0.25}{\includegraphics{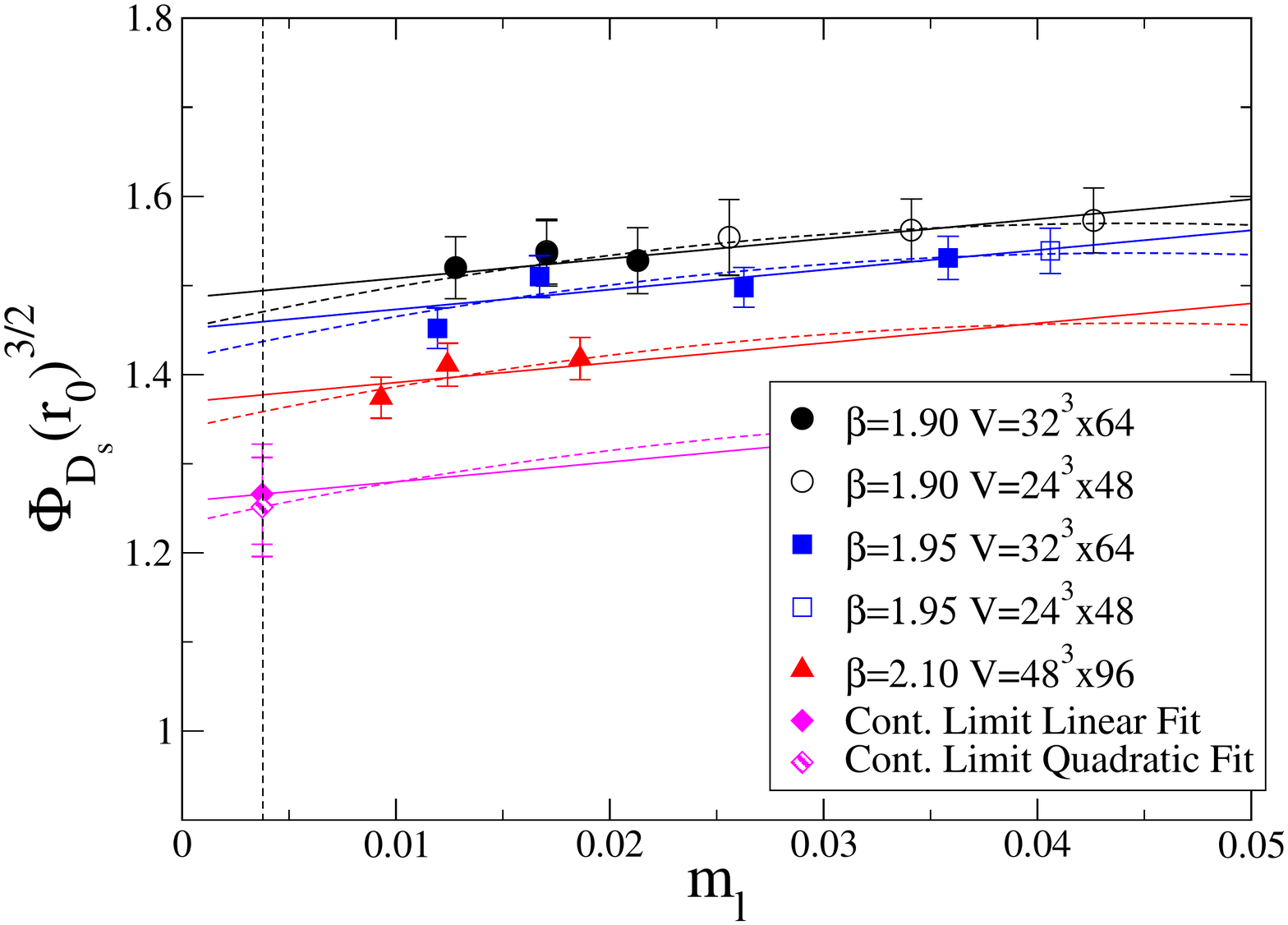}}
\scalebox{0.25}{\includegraphics{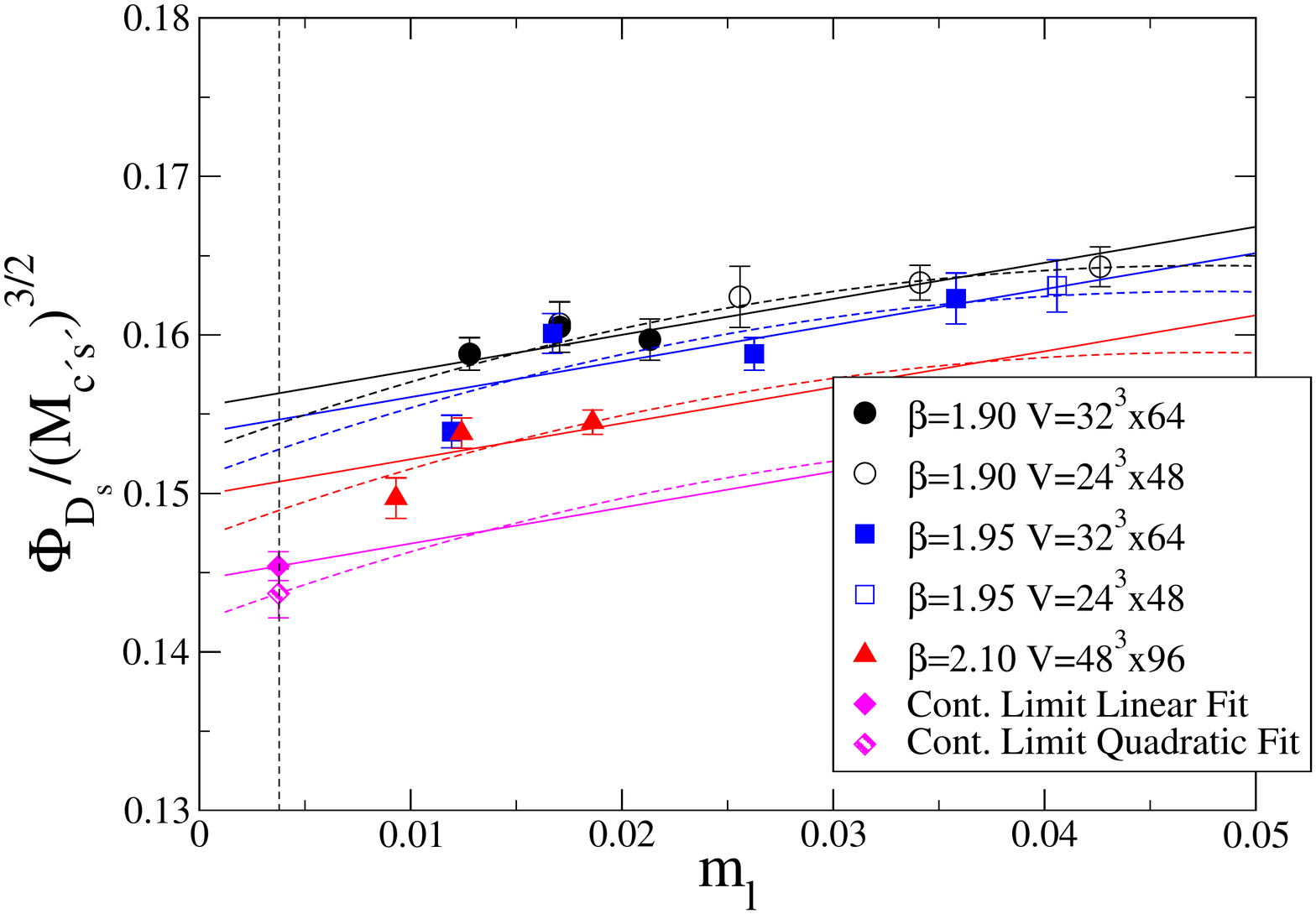}}
\caption{\it Chiral and continuum extrapolation of $r_0^{3/2} \Phi_{D_s}$ (left) and $\Phi_{D_s}/M_{c^\prime s^\prime}^{3/2}$ (right) based on Eq.~(\ref{eq:phids}).}
\label{fig:phids}
\end{figure}

The chiral and continuum extrapolations of $\Phi _{D_s } r^{3/2} _0$ and $\Phi _{D_s } /M_{c^\prime s^\prime}^{3/2}$ are shown in Fig.~\ref{fig:phids}.
The systematic uncertainty associated with the chiral extrapolation has been studied using both a linear or a quadratic fit in $m_\ell$.

We have taken into account also the error induced by the uncertainties on the physical strange and charm quark masses, and we have included it in the stat+fit error.
As for FSE, the comparison of our lattice data at the same quark mass and different lattice volumes shows that its effect is well within the statistical uncertainty. 
All the results, including the ones obtained using the $Z_P$ values from the M1 or M2 methods \cite{ETM:2011aa}, have been combined to get our final result for $f_{D_s}$, namely
\begin{eqnarray}\label{eq:fdsresults}
f_{D_s} = 242.1(7.6)_{stat + fit} (1.4)_{Chiral} (2.9)_{Disc.} (0.3)_{Z_P} \mev = 242.1(8.3) \mev ~ . 
\end{eqnarray}
Our result obtained at $N_f = 2 + 1 + 1$ can be compared with the FLAG averages $f_{D_s} = 248 (6) \mev$ at $N_f = 2$ and $f_{D_s} = 248.6 (2.7) \mev$ at $N_f = 2 + 1$ \cite{FLAG}.

\begin{figure}[htb!]
\centering
\scalebox{0.3}{\includegraphics{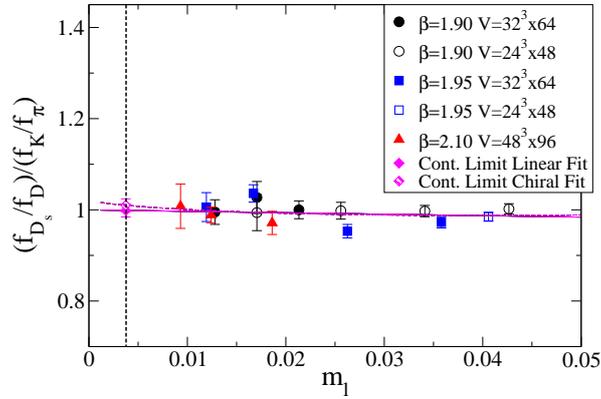}}
\caption{\it Chiral and continuum extrapolation of the double ratio $(f_{D_s}/f_D)/(f_K/f_\pi)$ using both the predictions (\ref{eq:fDsfD}) from ChPT and the polynomial expansion (\ref{eq:fDsfD_lin}) in the light quark mass $m_\ell$.}
\label{fig:doubleratio}
\end{figure}

The ratio $f_{D_s}/f_{D}$ can be calculated by analyzing the lattice data of the ratio $\Phi_{D_s} / \Phi_{D}$.
However, analyzing instead the double ratio $(f_{D_s}/f_D)/(f_K/f_\pi)$ increases the precision on the determination of $f_{Ds}/f_D$, because of the very mild dependence of the double ratio on $m_\ell$ \cite{Becirevic:2002mh}.
To fit the double ratio we combined the ChPT predictions for $f_\pi$ and $f_K$ with the HMChPT prediction for $\Phi_{D_s}/\Phi_{D}$, obtaining the following formula
\begin{equation} \label{eq:fDsfD}
\frac{{f_{D_s} / f_D }}{{f_K / f_\pi}} = P_1 \left[ {1 + P_2 m_\ell  + \left( {\frac{9}{4}\hat g^2  - \frac{1}{2}} \right)\xi _\ell \log \xi _\ell } \right] \frac {K^{FSE}_{f_\pi}} {K^{FSE}_{f_K}} \,,
\end{equation}
where for the coupling constant $\hat{g}$ we have considered the value $\hat{g} = 0.61 (7)$ \cite{Nakamura:2010zzi}, which among the presently available determinations of $\hat{g}$ maximizes the impact of the chiral log in Eq.~(\ref{eq:fDsfD}).
Notice that in Eq.~(\ref{eq:fDsfD}) any dependence of the double ratio on the lattice spacing is neglected, since such a dependence is not visible in the lattice data (see Fig.~\ref{fig:doubleratio}). We checked that by performing the same fit in $m_\ell$ without the data corresponding to the coarser lattice spacing (which corresponds roughly to keep half of the data) the same result for the double ratio is obtained.

Notice also in Eq.~(\ref{eq:fDsfD}) the presence of the FSE corrections for both $f_\pi$ and $f_K$. We described them using the CWW \cite{Colangelo:2010cu} and CDH \cite{CDH05} formulae, respectively, which reproduce the FSE observed in the data at the same light-quark mass and lattice spacing, but different lattice volumes.

An alternative fit with no chiral log was performed in order to evaluate the systematic error associated to chiral extrapolation, namely
\begin{equation} \label{eq:fDsfD_lin}
\frac{{f_{D_s } /f_D }}{{f_K /f_\pi  }} = P_1 \left( {1 + P_2 m_\ell} \right) \frac {K^{FSE}_{f_\pi}} {K^{FSE}_{f_K}} \,.
\end{equation}
The chiral extrapolation for the double ratio $(f_{D_s}/f_D)/(f_K/f_\pi)$, using both the ChPT fit (\ref{eq:fDsfD}) and the linear one (\ref{eq:fDsfD_lin}), is shown in Fig.~\ref{fig:doubleratio}. 

It can be seen that the two ans\"atze chosen for the chiral extrapolation provide compatible results for all pion masses within the statistical uncertainties. 
Our data have been extrapolated to $m_{ud}$ and therefore our results for $f_{D_s}/f_D$ correspond to the QCD isospin symmetric limit. 
%However, we estimated the impact of the isospin breaking effect to be below the per cent level hence small, in this case, compared to other sources of uncertainties.
Isospin breaking effects are estimated to be below the percent level and therefore small compared to other uncertainties.

The relevant source of systematic errors for the double ratio $(f_{D_s}/f_D)/(f_K/f_\pi)$ is the chiral extrapolation, while for $f_{D_s}/f_D$ also the uncertainty on $f_K/f_\pi$ has to be considered.
On the contrary, the errors on the physical strange and charm quark masses as well as the discretization errors and the uncertainty on the RC $Z_P$ contribute negligibly.

Our final results for $(f_{D_s}/f_D)/(f_K/f_\pi)$ and $f_{D_s}/f_D$ are
\begin{eqnarray}
\label{eq:doubleratioresults}
\frac{{f_{D_s } /f_D }}{{f_K /f_\pi  }} & = & 1.005(14)_{stat + fit} (6)_{Chiral} = 1.005(15) ~ , \\[2mm] 
\label{eq:fds/fdresults}
f_{D_s } /f_D  & = & 1.199(17)_{stat + fit} (7)_{Chiral} (16)_{f_K /f_\pi} = 1.199(25) ~ .
\end{eqnarray}
The latter one can be compared with the FLAG averages $f_{D_s } /f_D = 1.17 (5)$ at $N_f = 2$ and $f_{D_s } /f_D = 1.187 (12)$ at $N_f = 2 + 1$.
Notice the remarkable result for the double ratio given in Eq.~(\ref{eq:doubleratioresults}), which means that SU(3) breaking effects in the ratio of PS meson decay constants are the same in the light and charm sectors within a percent accuracy.

Finally we combined our results for $f_{D_s}$ and $f_{D_s}/f_D$ to obtain for $f_D$ the value
\begin{equation}
\label{eq:fdresults}
f_D  = 201.9(8.0) \mev ~ .
\end{equation}
The FLAG averages \cite{FLAG} are $f_D = 212 (8) \mev$ at $N_f = 2$ and $f_D = 209.2 (3.3) \mev$ at $N_f = 2+1$.

\section{Acknowledgements}

We acknowledge the CPU time provided by the PRACE Research Infrastructure under the project PRA027 ``QCD Simulations for Flavor Physics in the Standard Model and Beyond" on the JUGENE BG/P system at the J\"ulich SuperComputing Center (Germany), and by the agreement between INFN and CINECA under the specific initiative INFN-RM123.

\end{document}